\documentclass[aps,prl,reprint,nofootinbib]{revtex4-1}

\usepackage{amsmath}
\usepackage{amssymb}	
\usepackage{bbold}
\usepackage{bm} 
\usepackage{cancel} 
\usepackage{xcolor}
\usepackage[normalem]{ulem} 
\usepackage{changepage}
\usepackage{float}
\usepackage{mathrsfs}
\usepackage{url}
\usepackage{xfrac}
\usepackage{graphicx}
\usepackage{lipsum} 
\graphicspath{{/}}
\usepackage[font=small,labelfont=bf,justification=raggedright,format=plain,singlelinecheck=false]{caption} 

\newcommand{\comment}[1]{\ignorespaces} 

\renewcommand{\v}[1]{\ensuremath{\mathbf{#1}}} 
\newcommand{\gv}[1]{\ensuremath{\mbox{\boldmath$ #1 $}}} 
\newcommand{\pd}[2]{\frac{\partial #1}{\partial #2}} 
\providecommand{\e}[1]{\ensuremath{\times 10^{#1}}}
\DeclareMathAlphabet{\mathsfit}{\encodingdefault}{\sfdefault}{m}{n}
\SetMathAlphabet{\mathsfit}{bold}{\encodingdefault}{\sfdefault}{bx}{n}
\newcommand{\grad}[1]{\gv{\nabla} #1} 
\newcommand{\ws}[1]{\mathsf{#1}} 

\makeatletter
\newcommand*\bigcdot{\mathpalette\bigcdot@{.5}}
\newcommand*\bigcdot@[2]{\mathbin{\vcenter{\hbox{\scalebox{#2}{$\m@th#1\bullet$}}}}}
\makeatother

\usepackage{accents}

\usepackage{stackengine} 

\makeatletter 
\renewcommand*\env@matrix[1][*\c@MaxMatrixCols c]{%
  \hskip -\arraycolsep
  \let\@ifnextchar\new@ifnextchar
  \array{#1}}
\makeatother

\usepackage{pgfplots}
\usepgfplotslibrary{patchplots}
\usepgfplotslibrary{external}
\usepackage{tikz}
\usetikzlibrary{external}
\usetikzlibrary{calc}
\usetikzlibrary{arrows,shapes,backgrounds,arrows.meta,shapes.arrows,automata,quotes,intersections}
\usetikzlibrary{decorations.text}

\usetikzlibrary{decorations.pathreplacing}
\makeatletter
    \let\pgf@decorate@@brace@brace@code@old\pgf@decorate@@brace@brace@code
    \def\pgf@decorate@@brace@brace@code{
        \ifdim\pgfdecoratedremainingdistance<4\pgfdecorationsegmentamplitude
            \pgftransformxscale{\pgfdecoratedremainingdistance/4\pgfdecorationsegmentamplitude}
            \pgfdecoratedremainingdistance=4\pgfdecorationsegmentamplitude
        \fi
        \pgf@decorate@@brace@brace@code@old
    }
\makeatother

\newcommand*\mL{\mathcal{L}}

\newcommand*\mOne{\mathbb{1}}
\newcommand*\mZero{\mathbb{0}}
\newcommand*\vphi{\varphi}
\newcommand*\tphi{\tilde{\phi}}

\DeclareMathAlphabet{\mathpzc}{OT1}{pzc}{m}{it}

\let\emptyset\varnothing

\usepackage{bbding}
\DeclareFontFamily{U}{MnSymbolC}{}
\DeclareSymbolFont{MnSyC}{U}{MnSymbolC}{m}{n}
\DeclareMathSymbol{\diamondplus}{\mathbin}{MnSyC}{"7C}
\DeclareMathSymbol{\diamonddot}{\mathbin}{MnSyC}{"7E}
\DeclareFontShape{U}{MnSymbolC}{m}{n}{
    <-6>  MnSymbolC5
   <6-7>  MnSymbolC6
   <7-8>  MnSymbolC7
   <8-9>  MnSymbolC8
   <9-10> MnSymbolC9
  <10-12> MnSymbolC10
  <12->   MnSymbolC12}{}

\makeatletter
\newcommand*{\defeq}{\mathrel{\rlap{%
                     \raisebox{0.3ex}{$\m@th\cdot$}}%
                     \raisebox{-0.3ex}{$\m@th\cdot$}}%
                     =}
\newcommand*{\eqdef}{=\mathrel{\rlap{%
                     \raisebox{0.3ex}{$\m@th\cdot$}}%
                     \raisebox{-0.3ex}{$\m@th\cdot$}}%
                     }
\makeatother

\usepackage{mathtools}

\usepackage{stmaryrd}

\usepackage{enumitem}

\usepackage{diagrams}
\usetikzlibrary{cd}

\definecolor{light-gray}{gray}{.85}
\newsavebox{\songboxbox}

\usepackage[most]{tcolorbox}
\newtcbox{\MyBox}[1][red]{on line, size=tight, boxsep=1pt, colframe=#1!50!black, colback=#1!10!white}

\usepackage[framemethod=TikZ]{mdframed}
\mdfsetup{%
    outerlinewidth=1pt,
    innertopmargin=6pt,
    innerbottommargin=6pt,
    skipabove=10pt,
    skipbelow=10pt,
    backgroundcolor=black!10,
    roundcorner=20pt
}

\DeclareFontFamily{U}{matha}{\hyphenchar\font45}
\DeclareFontShape{U}{matha}{m}{n}{
      <5> <6> <7> <8> <9> <10> gen * matha
      <10.95> matha10 <12> <14.4> <17.28> <20.74> <24.88> matha12
      }{}
\makeatletter
\newcommand{\blandor}[1]{\mathbin{\@blandor{#1}}}
\newcommand{\@blandor}[1]{\mathchoice
  {\@@blandor{#1}{\tf@size}}
  {\@@blandor{#1}{\tf@size}}
  {\@@blandor{#1}{\sf@size}}
  {\@@blandor{#1}{\ssf@size}}
}
\newcommand{\@@blandor}[2]{%
    \raisebox{.1ex}{\rotatebox[origin=c]{#1}{%
      \fontsize{#2}{#2}\usefont{U}{matha}{m}{n}\symbol{\string"CE}}}%
}
\makeatother

\usepackage{slashed} 
\newcommand{\cmmnt}[1]{\ignorespaces} 

\usepackage{environ}
\NewEnviron{eqn}{
\begin{align}\begin{split}
\BODY
\end{split}\end{align}
}

\begin{document}


\title{Lifting Spacetime's Poincar\'e Symmetries} 



\author{Alexander S. Glasser}
\affiliation{Department of Astrophysical Sciences, Princeton University, Princeton, New Jersey 08540}
\author{Hong Qin}
\affiliation{Department of Astrophysical Sciences, Princeton University, Princeton, New Jersey 08540}


\date{\today}
\begin{abstract}
In the following work, we pedagogically develop \emph{5-vector theory}, an evolution of scalar field theory that provides a stepping stone toward a Poincar\'e-invariant lattice gauge theory.  Defining a continuous flat background via the four-dimensional Cartesian coordinates $\{x^a\}$, we `lift' the generators of the Poincar\'e group so that they transform only the fields existing upon $\{x^a\}$, and do not transform the background $\{x^a\}$ itself.  To facilitate this effort, we develop a non-unitary particle representation of the Poincar\'e group, replacing the classical scalar field with a 5-vector matter field.  We further augment the vierbein into a new $5\times5$ \emph{f\"unfbein}, which `solders' the 5-vector field to $\{x^a\}$.  In so doing, we form a new intuition for the Poincar\'e symmetries of scalar field theory.  This effort recasts `spacetime data', stored in the derivatives of the scalar field, as `matter field data', stored in the 5-vector field itself.  We discuss the physical implications of this `Poincar\'e lift', including the readmittance of an absolute reference frame into relativistic field theory.  In a companion paper \cite{glasser_discrete5vectortheory}, we demonstrate that this theoretical development, here construed in a continuous universe, enables the description of a discrete universe that preserves the 10 infinitesimal Poincar\'e symmetries and their conservation laws.
\end{abstract}

\pacs{}

\maketitle 


\section{Introduction}

The Poincar\'e group sharply contrasts with the ${SU(3)\times SU(2)\times U(1)}$ symmetry group of the Standard Model, in that it is understood to act not only on particles of the Standard Model, but also on the spacetime manifold itself.  This foundational assumption of the Standard Model sets gravity apart from the other three fundamental forces.

In the present work, we reconstrue the symmetries of the Poincar\'e group, not as `horizontal' symmetries acting on the spacetime manifold, but as `lifted' symmetries acting on a physical theory's `vertical' space.  We develop \emph{5-vector theory}, an evolution of classical scalar field theory whose Poincar\'e symmetries are elevated to act solely on its vertical matter and solder fields---and whose dynamics are nevertheless equivalent to those of the scalar field.

To accommodate such a modified group action, we reimagine both the matter field of scalar field theory as well as the `background canvas' on which scalar field theory is constructed.  Rather than embedding our physical model in a spacetime whose coordinates transform with translations and Lorentz transformations, we imagine a `static' four-dimensional background, with Cartesian coordinates $\{x^a\}$ and metric $\eta_{ab}$ of signature ($-$$+$$+$$+$).  This Cartesian\footnote{Although they have a defined metric, the coordinates $\{x^a\}_{a\in\{0,1,2,3\}}$ are labeled as \emph{Cartesian}---and not \emph{Euclidean} or \emph{Minkowskian}---to emphasize their rigidity.  Their discrete counterpart---the `integer lattice'---will not have \emph{any} infinitesimal symmetries.\label{whyCartesian}} spacetime is not to be viewed as a Poincar\'e-set---in particular, it does not transform under Poincar\'e transformations.  In this sense, the coordinate $x^0$ represents an absolute time and $\v{x}$ an absolute space---together comprising an absolute reference frame of what we will show to be, nonetheless, a fully relativistic physical model.

On this background, we replace the familiar scalar field $\phi(x^b)$ with a \emph{5-vector} matter field, and we augment the $4\times4$ vierbein matrix $e_\mu^{~a}(x^b)$ into a solder field\footnote{We note that our use of the term \emph{solder field} is exceptional; in treatments of reductive Cartan geometry (e.g. \cite{wise_macdowell-mansouri_2010}), the \emph{soldering form} or \emph{coframe field} is a $\mathfrak{g}/\mathfrak{h}$-valued horizontal 1-form that shares the dimension of its base manifold.  Because we augment the vierbein into the larger f\"unfbein, however, we find it simplest to regard both $e_\mu^{~a}(x^b)$ and $\v{e}(x^b)$ as matrix-valued 0-forms.} we call the \emph{f\"unfbein}, a $5\times5$ matrix at each point $x^b$, as follows:
\begin{eqn}
\phi(x^b)&\rightarrow\gv{\phi}(x^b)\defeq\left[\begin{matrix}[l]\phi^\mu\\~\phi\end{matrix}\right](x^b)\\
\vspace{10pt}\\
e_\mu^{~a}(x^b)&\rightarrow\v{e}(x^b)\defeq\left[\begin{matrix}e_\mu^{~a}&\v{0}\\e_\mu&1\end{matrix}\right](x^b).
\label{new5Fields}
\end{eqn}
In a manner that shall be made clear, the f\"unfbein serves as the `hinges' that bind, or solder, the vertical matter field $\gv{\phi}$ to the horizontal Cartesian background; we shall see that $\{x^a\}$ itself is independent of Poincar\'e transformations, and that the transformed f\"unfbein provides the mapping from this fixed background to the transformable matter field.  It is for this reason that both Poincar\'e (Greek) indices and Cartesian (Latin) indices appear in the f\"unfbein.

The 5-vector and f\"unfbein serve as the targets of Poincar\'e transformations---as the Poincar\'e-sets---of our new field theory.  As we will show, in establishing a more detailed matter field $\gv{\phi}$, we essentially transfer the data stored in spacetime---via spacetime derivatives $\partial^\mu\phi$ of the scalar field---to data stored in the $\phi^\mu$ components of the matter field itself.  As further demonstrated in a companion paper \cite{glasser_discrete5vectortheory}, this data-transfer is crucial---it `unburdens' the Cartesian canvas of our physical theory, affording its discretization without sacrificing Poincar\'e symmetry.

In the following work, we motivate this evolution of scalar field theory.  We apply at length the variational technology of \cite{olver_textbook_1993}, and progressively introduce modified forms of the scalar field Lagrangian until our 5-vector theory is discovered.  We solve for the symmetries and Poincar\'e currents---the linear and angular energy-momentum tensors---of our theory, and we discuss the physical implications of this `Poincar\'e lift'.

\section{Real Scalar Field Theory}

We begin with a review of the Lagrangian for a real scalar field with an arbitrary potential $V(\phi)$ in ($-$$+$$+$$+$) Minkowski spacetime:
\begin{eqn}
\mL\defeq-\frac{1}{2}\partial_\mu\phi\partial^\mu\phi-V(\phi).
\label{phiLagrangian}
\end{eqn}
Applying the Euler operator\footnote{
An introduction of notation is helpful here.  For a system of $M$ independent variables $\{x^i\}$ and $N$ dependent variables $\left\{u^\ell\right\}$, we denote a \emph{$(k\geq0)$-order multi-index} $J$ by $J\equiv(j_1,\dots,j_k)$, where ${1\leq j_i\leq M}$.  We let $\#J$ denote the order (i.e., length) of the multi-index $J$, where any repetitions of indices are to be double-counted.  Accordingly, $u_J$ represents a partial derivative taken with respect to $(x^{j_1},\dots,x^{j_k})$.  For example, ${u_{ii}\equiv\partial^2u/\partial x^i\partial x^i}$ has $\#J=2$.
$\ws{D}_i$ denotes a \emph{total derivative}, i.e.:
\begin{equation*}
\ws{D}_iP\defeq\pd{P}{x^i}+\sum\limits_{\ell=1}^N\sum\limits_{\#J\geq0}u_{J,i}^\ell\pd{P}{u^\ell_J}
\end{equation*}
and $\ws{D}_J\equiv\ws{D}_{j_1}\cdots\ws{D}_{j_k}$.  We let $(-\ws{D})_J$ denote a total derivative with negative signs included for each index.  For example:
\begin{equation*}
(-\ws{D})_{xyz}u=(-\ws{D}_x)(-\ws{D}_y)(-\ws{D}_z)u=-\pd{^3u}{x\partial y\partial z}\equiv-u_{xyz}.
\end{equation*}
As is conventional, we will sometimes relax our notation and denote the total and partial derivatives by the same symbol: $\partial_i$.
}
for $\phi$---
\begin{eqn}
\ws{E}_\phi\defeq&\sum\limits_J(-\ws{D})_J\pd{}{\left(\phi_J\right)}\\
=&\partial_\phi-\ws{D}_\mu\partial_{(\partial_\mu\phi)}+\cdots
\end{eqn}
---we derive the following equation of motion (EOM):
\begin{eqn}
0=\ws{E}_\phi(\mL)=-V'(\phi)+\partial_\mu\partial^\mu\phi.
\end{eqn}

Let us review the usual Poincar\'e symmetries associated with this Lagrangian---in particular its translation symmetry generators $P_\alpha$ and Lorentz symmetry generators $M_{\alpha\beta}$, for ${\alpha\neq\beta\in\{t,x,y,z\}}$:
\begin{eqn}
P_\alpha&\defeq\partial_\alpha\\
M_{\alpha\beta}&\defeq x_\alpha\partial_\beta-x_\beta\partial_\alpha
\label{PoincareSymmetries}
\end{eqn}
where ${\partial_\mu\equiv\partial/\partial x^\mu}$ and ${x_\mu=\eta_{\mu\nu}x^\nu}$.  (For now, we employ the most familiar setting for field theories---a flat, four-dimensional spacetime labeled by Poincar\'e-transformable coordinates $x^\mu$.)  These symmetries are called \emph{spacetime symmetries} because they operate on spacetime itself---that is, on the \emph{independent} or \emph{horizontal} variables of the theory.

We denote the $n^\text{th}$\emph{-order jet space} ${\ws{Jet}(n)}$ of scalar theory---comprised of (i) spacetime's independent coordinate variables $x^\mu$; (ii) the dependent variable $\phi(x^\mu)$; and (iii) the derivatives of the dependent variable up to order $n$: $\partial_{\mu_1\cdots \mu_n}\phi(x^\mu)$---as follows:
\begin{eqn}
\ws{Jet}(n)&\equiv X\times U^{(n)}\\
&\equiv X\times U\times U^1\times\cdots\times U^n
\end{eqn}
for $x^\mu\in X$, $\phi\in U$, $\partial_\mu\phi\in U^1$ and so on.  $X$ is referred to as the \emph{horizontal subspace} of $\ws{Jet}(n)$, and $U^{(n)}$ as the \emph{vertical subspace} of $\ws{Jet}(n)$.  Correspondingly, the jet space differential operator $\partial_\mu$ is referred to as a \emph{horizontal vector field}, while ${\partial_\phi+\partial_{(\partial_\mu\phi)}}$, for example, is referred to as a \emph{vertical vector field}.

Generalizing for the moment to an \emph{arbitrary} ${0^\text{th}\text{-order}}$ jet space ${\ws{Jet}(0)=X\times U}$, comprised of ${M=\ws{dim}(X)}$ horizontal and ${N=\ws{dim}(U)}$ vertical variables, we briefly review the elements of \cite{olver_textbook_1993} essential to our study.  Following \cite{olver_textbook_1993} Eq.~(5.1), we define a \emph{generalized vector field} $\v{v}$ on $\ws{Jet}(0)$:
\begin{eqn}
\v{v}=\sum\limits_{i=1}^M\zeta^i[u]\pd{}{x^i}+\sum\limits_{\ell=1}^N\gamma^\ell[u]\pd{}{u^\ell}
\label{defineV}
\end{eqn}
where $\zeta^i$ and $\gamma^\ell$ are arbitrary smooth functions, and where ${[u]\equiv\left(x,u^{(n)}\right)}$ denotes their dependence on any variables of $\ws{Jet}(n)$, such that:
\begin{eqn}
[u]\defeq\Big(x^i\in X,u^\ell\in U,u^\ell_{x^i}\in U^1,\dots,u^\ell_{x^{i_1}\cdots x^{i_n}}\in U^n\Big).
\end{eqn}
$\v{v}$ is understood to be the generator of a smooth transformation of the variables in $\ws{Jet}(0)$.  There is a unique extension of $\v{v}$ from ${X\times U}$ to ${X\times U^{(n)}}$ that self-consistently specifies the flow of the vertical `derivative subspace' $U^{[1,n]}\subset\ws{Jet}(n)$, given the flow of $\ws{Jet}(0)$ along $\v{v}$.  This extension is referred to as the vector field's \emph{prolongation} and is given by:
\begin{eqn}
\ws{pr}[\v{v}]=\sum\limits_{i=1}^M\zeta^i[u]\pd{}{x^i}~+~\smashoperator{\sum\limits_{\substack{\ell\in[1,N]\\\#J\in[0,n]}}}\Omega^J_\ell[u]\pd{}{u^\ell_J}
\label{prolongationOfV}
\end{eqn}
where
\begin{eqn}
\Omega^J_\ell[u]\defeq\ws{D}_J\left(\gamma^\ell[u]-\sum\limits_{i=1}^Mu^\ell_i\zeta^i[u]\right)+\sum\limits_{i=1}^Mu^\ell_{J,i}\zeta^i[u]
\end{eqn}
and where the sum over ${\#J\in[0,n]}$ indicates a sum over all multi-indices of length ${0\leq\#J\leq n}$.  (Note that ${\ws{pr}[\v{v}]=\v{v}}$ when restricted to its ${\#J=0}$ terms.)  A jet space vector field $\v{v}$ is defined to act on (i.e., infinitesimally transform) any Lagrangian $\mL[u]$ via its prolongation: $\ws{pr}[\v{v}](\mL)$.

Still following \cite{olver_textbook_1993}, we define the \emph{evolutionary representative} of $\v{v}$ by the following vertical vector field $\v{v}_Q$:
\begin{eqn}
\v{v}_Q=\smashoperator{\sum\limits_{\substack{\ell\in[1,N]}}}Q_\ell[u]\pd{}{u^\ell}
\label{evolRep}
\end{eqn}
where the \emph{characteristics} $Q_\ell[u]$ of $\v{v}_Q$ are given by
\begin{eqn}
Q_\ell[u]\defeq\gamma^\ell[u]-\sum\limits_{i=1}^Mu^\ell_i\zeta^i[u].
\label{charOfEvol}
\end{eqn}
Following Eq.~(\ref{prolongationOfV}), we calculate the prolongation of $\v{v}_Q$ as follows:
\begin{eqn}
\ws{pr}[\v{v}_Q]=\smashoperator{\sum\limits_{\substack{\ell\in[1,N]\\\#J\in[0,n]}}}\Big(\ws{D}_JQ_\ell[u]\Big)\pd{}{u^\ell_J}.
\label{prolongOfEvol}
\end{eqn}

We define a \emph{variational symmetry} of the Lagrangian $\mL$ as a generalized vector field $\v{v}$ for which there exists an $M$-tuple ${B=(B^1[u],\dots,B^M[u])}$ such that either of the following equivalent conditions hold:\footnote{We interchangeably employ the notations:
\begin{equation*}
\ws{Div}P\equiv\ws{Div}_i P^i\equiv\sum\limits_{i=1}^M\ws{D}_iP^i.
\end{equation*}
}
\begin{eqn}
\ws{pr}[\v{v}](\mL)+\mL~\ws{Div}_i\zeta^i&=\ws{Div}_i(B^i+\mL\zeta^i)\\
\ws{pr}[\v{v}_Q](\mL)&=\ws{Div}B.
\label{defineVarSymm}
\end{eqn}
As proven in \cite{olver_textbook_1993} Proposition 5.52, $\v{v}$ is a variational symmetry of $\mL[u]$ if and only if $\v{v}_Q$ is.  (We note that $\v{v}_Q$ is its own evolutionary representative.)

Noether's theorem establishes a one-to-one correspondence between the (equivalence classes of) variational symmetries of a Lagrangian and (equivalence classes of) its conservation laws.\footnote{We refer the reader to \cite{olver_textbook_1993} pp. 264, 292 for a discussion of \emph{trivial} symmetries and conservations laws, and the \emph{equivalence classes} they generate.\label{footnoteEquivTrivial}}  Because $\v{v}$ and $\v{v}_Q$ belong to the same equivalence class, the Noether procedure for $\v{v}_Q$ discovers conservation laws equivalent to those of $\v{v}$.

Returning now to our present scalar theory, therefore, we first find the evolutionary representatives ${P_\alpha}_Q$ and ${M_{\alpha\beta}}_Q$ of the Poincar\'e symmetries by applying the framework of Eqs.~(\ref{defineV})-(\ref{prolongOfEvol}) to the Poincar\'e generators of Eq.~(\ref{PoincareSymmetries}).  We derive their prolongations $\ws{pr}[{P_\alpha}_Q]$ and $\ws{pr}[{M_{\alpha\beta}}_Q]$ to first order, which suffices because higher order derivatives do not appear in the Lagrangian $\mL$ of Eq.~(\ref{phiLagrangian}).  We find:
\begin{eqn}
\ws{pr}\left[{P_\alpha}_Q\right]&=-\pd{\phi}{x^\alpha}\pd{}{\phi}-\ws{D}_\mu\left[\pd{\phi}{x^\alpha}\right]\pd{}{(\partial_\mu\phi)}\\
\ws{pr}\left[{M_{\alpha\beta}}_Q\right]&=-\left(x_\alpha\pd{\phi}{x^\beta}-x_\beta\pd{\phi}{x^\alpha}\right)\pd{}{\phi}\\
&-\ws{D}_\mu\left[x_\alpha\pd{\phi}{x^\beta}-x_\beta\pd{\phi}{x^\alpha}\right]\pd{}{(\partial_\mu\phi)}.
\label{verticalSymms}
\end{eqn}
Applying these prolonged vector fields to $\mL$, we calculate as follows:
\begin{eqn}
\ws{pr}\left[{P_\alpha}_Q\right](\mL)&=-\ws{D}_\mu(\delta^\mu_\alpha\mL)\\
\ws{pr}\left[{M_{\alpha\beta}}_Q\right](\mL)&=-\ws{D}_\mu\left[\left(x_\alpha\delta^\mu_\beta-x_\beta\delta^\mu_\alpha\right)\mL\right].
\label{phiTheoryDivergencesVQ}
\end{eqn}
It is immediately seen that
\begin{eqn}
\ws{pr}\left[\v{v}_Q\right](\mL)=-\ws{Div}\left(\mL\zeta\right)
\label{resultOfVQ}
\end{eqn}
for each of the Poincar\'e symmetries ${\v{v}_Q={P_\alpha}_Q}$ and ${\v{v}_Q={M_{\alpha\beta}}_Q}$, where $\zeta$ refers to the horizontal coefficients of Eq.~(\ref{defineV}), as defined in Eq.~(\ref{PoincareSymmetries}).

Eq.~(\ref{phiTheoryDivergencesVQ}) demonstrates that ${P_\alpha}$ and ${M_{\alpha\beta}}$---and their evolutionary representatives---are variational symmetries of $\mL$, as defined in Eq.~(\ref{defineVarSymm}).  By Noether's theorem, then, we are guaranteed to find conservation laws associated to each of these symmetries, as we now do.

We carry out the Noether procedure as specified in \cite{olver_textbook_1993} Proposition 5.98 to explicitly solve for these conservation laws.  To do so, we introduce the higher Euler operators $\ws{E}^J_{u^\ell}$, whose purpose is to facilitate the following `integration by parts' for \emph{any} vertical vector field $\v{v}_Q$ with characteristics $Q_\ell$, as notated in Eq.~(\ref{evolRep}):
\begin{eqn}
\ws{pr}[\v{v}_Q](P)=\sum\limits_{\ell=1}^N\sum\limits_{\#J\geq0}\ws{D}_J\left(Q_\ell\cdot\ws{E}^J_{u^\ell}(P)\right)
\label{higherEulerDef}
\end{eqn}
for any $P$.  Eq.~(\ref{higherEulerDef}) is definitional, as it uniquely determines the form of these operators, as follows:\footnote{We let ${\left(\begin{smallmatrix}I\\J\end{smallmatrix}\right)\equiv I!/[J!(I\backslash J)!]}$ when ${J\subseteq I}$, and $0$ otherwise.  We define $I!=(\tilde{i}_1!\cdots\tilde{i}_M!)$, where $\tilde{i}_k$ denotes the number of occurrences of the integer $k$ in multi-index $I$.  $I\backslash J$ denotes the set difference of multi-indices, with repeated indices treated as distinct elements of the set, as they are in ${J\subseteq I}$.}
\begin{eqn}
\ws{E}^J_{u^\ell}\defeq\sum\limits_{I\supseteq J}\left(\begin{matrix}I\\J\end{matrix}\right)(-\ws{D})_{I\backslash J}\pd{}{u^\ell_I}.
\end{eqn}
We note that ${\ws{E}^J_{u^\ell}\equiv\ws{E}_{u^\ell}}$ is the conventional Euler operator for ${J=\emptyset}$.

For ${P=\mL}$ a Lagrangian, one observes that the right hand side of Eq.~(\ref{higherEulerDef}) splits $\ws{pr}[\v{v}_Q](\mL)$ into a divergence and a term that vanishes along solutions ${\ws{E}_{u^\ell}(\mL)=0}$ of our system---that is, \emph{on shell}:
\begin{eqn}
\ws{pr}\left[\v{v}_Q\right](\mL)=\left(\sum\limits_{\ell=1}^NQ_\ell\cdot\ws{E}_{u^\ell}(\mL)\right)+\ws{Div}A
\label{prQ2QEandA}
\end{eqn}
where the $M$-tuple ${A=(A^1[u],\dots,A^M[u])}$ is given by
\begin{eqn}
A^k&\defeq\sum\limits_{\ell=1}^N\sum\limits_{\#I\geq0}\frac{\tilde{i}_k+1}{\#I+1}\ws{D}_I\left[Q_\ell\ws{E}_{u^\ell}^{I,k}(\mL)\right].
\label{defFourTuple}
\end{eqn}

Combining the observation of Eq.~(\ref{prQ2QEandA}) with the second relation of Eq.~(\ref{defineVarSymm}), we see that a variational symmetry $\v{v}_Q$ yields the following conservation law on shell:
\begin{eqn}
\ws{Div}\left(A-B\right)=0.
\label{conciseConsLawForVariational}
\end{eqn}

We may now carry out this Noether procedure for our particular first-order scalar field Lagrangian.  We need only find the higher Euler operator for multi-index ${J=(x^\mu)}$ and dependent variable ${u^\ell=\phi}$, as follows:
\begin{eqn}
\ws{E}_\phi^\mu(\mL)=\pd{\mL}{(\partial_\mu\phi)}=-\partial^\mu\phi.
\end{eqn}
For our Poincar\'e symetries ${P_\alpha}_Q$ and ${M_{\alpha\beta}}_Q$, therefore:
\begin{eqn}
A^\mu_{P_\alpha}&=(\partial_\alpha\phi)(\partial^\mu\phi)\\
A^\mu_{M_{\alpha\beta}}&=\left(x_\alpha\pd{\phi}{x^\beta}-x_\beta\pd{\phi}{x^\alpha}\right)(\partial^\mu\phi).
\label{TuplesForScalarTheory}
\end{eqn}
We correspondingly substitute $A$ from Eq.~(\ref{TuplesForScalarTheory}) and ${B=-\mL\zeta}$ from Eq.~(\ref{resultOfVQ}) for each respective symmetry into Eq.~(\ref{conciseConsLawForVariational}) to solve for our 10 conservation laws:
\begin{eqn}
0&=\ws{D}_\mu\Bigg[\partial^\mu\phi\partial^\alpha\phi+\eta^{\mu\alpha}\mL\Bigg]\eqdef\ws{D}_\mu T^{\mu\alpha}\\
0&=\ws{D}_\mu\Bigg[x^\alpha T^{\mu\beta}-x^\beta T^{\mu\alpha}\Bigg]\eqdef\ws{D}_\mu L^{\mu\alpha\beta}.
\label{twoConsLawsOfOriginalTheory}
\end{eqn}

We have thus found that the formal manipulations of \cite{olver_textbook_1993} recover the familiar conservation laws of scalar field theory.

\section{Real `Scalar + 4-Vector' Field Theory}

We now explore the Lagrangian of a new field theory that replicates the dynamics of the familiar scalar theory described above.  In part, we are inspired toward the following Lagrangian by the Goldstone model of \cite{kibble_symmetry_1967}.  We define:
\begin{eqn}
\mL&\defeq\frac{1}{2}\phi^\mu\phi_\mu-\frac{1}{2}\phi^\mu\partial_\mu\phi+\frac{1}{2}\phi\partial_\mu\phi^\mu-V(\phi).
\label{ScalarPlus4VectorLagrangian}
\end{eqn}
This Lagrangian has five dynamical variables, in the form of a Lorentz 4-vector $\phi^\mu$ and a scalar $\phi$.  In this flat-spacetime theory, (still possessing familiar `Poincar\'e-deformable' coordinates), we raise and lower Greek indices with the Minkowski metric $\eta_{\mu\nu}$ of signature $($$-$$+$$+$$+$$)$---for example: ${\phi_\mu=\eta_{\mu\nu}\phi^\nu}$ and ${\partial^\mu=\eta^{\mu\nu}\partial_\nu}$.

We again apply Euler operators to derive the following five EOM:
\begin{eqn}
0&=\ws{E}_{\phi}(\mL)=-V'(\phi)+\partial_\mu\phi^\mu\\
0&=\ws{E}_{\phi^\sigma}(\mL)=\phi_\sigma-\partial_\sigma\phi.
\label{RealPlusFourVectorEOM}
\end{eqn}
Upon combining these EOM, we see that our new Lagrangian replicates the dynamics of the familiar scalar field Lagrangian.  Unlike the scalar field Lagrangian, however, our new Lagrangian produces this behavior with coupled first-order EOM.

The latter EOM of Eq.~(\ref{RealPlusFourVectorEOM}) suggests that $\phi_\mu$ functions as the spacetime derivatives of the scalar field.  In this respect, the theory's variables are reminiscent of a Hamiltonian system, with first-order EOM given by
\begin{eqn}
\left[\begin{matrix}\dot{\v{x}}\\\dot{\v{p}}\end{matrix}\right]=\left[\begin{matrix}\v{p}\\-\grad{V(\v{x})}\end{matrix}\right].
\end{eqn}
Still, the EOM's differences from the Hamiltonian formulation---including its democratic, relativistic treatment of space and time---are perhaps more important.

We now repeat the exercise of the prior section and calculate in the formalism of \cite{olver_textbook_1993} the symmetries and conservation laws of the Lagrangian in Eq.~(\ref{ScalarPlus4VectorLagrangian}).  The dependent $\phi^\mu$ variables modify much of the analysis.

We begin with the same \emph{translation} symmetry generator $P_\alpha$ of Eq.~(\ref{PoincareSymmetries}), but we must modify $M_{\alpha\beta}$ to account for our new $\phi^\mu$ field.  After all, as indicated by Eq.~(\ref{RealPlusFourVectorEOM}), $\phi_\mu$ must transform as a spacetime derivative.  We therefore set:
\begin{eqn}
P_\alpha&\defeq\partial_\alpha\\
M_{\alpha\beta}&\defeq x_\alpha\partial_\beta-x_\beta\partial_\alpha+\phi_\alpha\partial_{\phi^\beta}-\phi_\beta\partial_{\phi^\alpha}
\label{FourVectorTheoryGenerators}
\end{eqn}
Calculating and applying prolongations of these symmetries to $\mL$, we find that:
\begin{eqn}
\ws{pr}\left[P_\alpha\right](\mL)&=0\\
\ws{pr}\left[M_{\alpha\beta}\right](\mL)&=0.
\end{eqn}
$\mL$ of Eq.~(\ref{ScalarPlus4VectorLagrangian}) is therefore invariant under transformation by the vector fields $P_\alpha$ and $M_{\alpha\beta}$ of Eq.~(\ref{FourVectorTheoryGenerators}).  According to the first relation of Eq.~(\ref{defineVarSymm}), setting ${B^i=-\mL\zeta^i}$ and noting that ${\ws{D}_i\zeta^i=0}$ for each symmetry, respectively, $P_\alpha$ and $M_{\alpha\beta}$ are indeed variational symmetries of $\mL$.

To find `scalar + 4-vector' theory's associated conservation laws, therefore, we may again solve for the prolonged evolutionary representatives of the new Poincar\'e symmetries of Eq.~(\ref{FourVectorTheoryGenerators}), up to first order:
\begin{eqn}
\ws{pr}\left[{P_\alpha}_Q\right]&=-\pd{\phi}{x^\alpha}\pd{}{\phi}-\pd{\phi^\sigma}{x^\alpha}\pd{}{\phi^\sigma}\\
&-\ws{D}_\mu\left[\pd{\phi}{x^\alpha}\right]\pd{}{(\partial_\mu\phi)}-\ws{D}_\mu\left[\pd{\phi^\sigma}{x^\alpha}\right]\pd{}{(\partial_\mu\phi^\sigma)}\\
\ws{pr}\left[{M_{\alpha\beta}}_Q\right]&=-\left(x_\alpha\pd{\phi}{x^\beta}-x_\beta\pd{\phi}{x^\alpha}\right)\pd{}{\phi}\\
&-\left(x_\alpha\pd{\phi^\sigma}{x^\beta}-x_\beta\pd{\phi^\sigma}{x^\alpha}\right)\pd{}{\phi^\sigma}\\
&+\phi_\alpha\pd{}{\phi^\beta}-\phi_\beta\pd{}{\phi^\alpha}\\
&-\ws{D}_\mu\left[x_\alpha\pd{\phi}{x^\beta}-x_\beta\pd{\phi}{x^\alpha}\right]\pd{}{(\partial_\mu\phi)}\\
&-\ws{D}_\mu\left[x_\alpha\pd{\phi^\sigma}{x^\beta}-x_\beta\pd{\phi^\sigma}{x^\alpha}\right]\pd{}{(\partial_\mu\phi^\sigma)}\\
&+\left[\pd{\phi_\alpha}{x^\mu}\pd{}{(\partial_\mu\phi^\beta)}-\pd{\phi_\beta}{x^\mu}\pd{}{(\partial_\mu\phi^\alpha)}\right].
\label{prolongedVertFiveVectorSymmetries}
\end{eqn}

Applying these to the Lagrangian of Eq.~(\ref{ScalarPlus4VectorLagrangian}), we find that ${P_\alpha}_Q$ and ${M_{\alpha\beta}}_Q$ again satisfy Eq.~(\ref{resultOfVQ}):
\begin{eqn}
\ws{pr}\left[{P_\alpha}_Q\right](\mL)&=-\ws{D}_\mu(\delta^\mu_\alpha\mL)\\
\ws{pr}\left[{M_{\alpha\beta}}_Q\right](\mL)&=-\ws{D}_\mu\left[\left(x_\alpha\delta^\mu_\beta-x_\beta\delta^\mu_\alpha\right)\mL\right].
\label{resultOfVQ2}
\end{eqn}
Eq.~(\ref{resultOfVQ2}) demonstrates that ${P_\alpha}_Q$ and ${M_{\alpha\beta}}_Q$---prolonged in Eq.~(\ref{prolongedVertFiveVectorSymmetries})---are also variational symmetries of the Lagrangian in Eq.~(\ref{ScalarPlus4VectorLagrangian}), as they had to be.  We can therefore solve for their associated currents via the Noether procedure of Eqs.~(\ref{higherEulerDef})-(\ref{conciseConsLawForVariational}).

To solve as before for our conserved currents, we first derive our new system's higher Euler operators:
\begin{eqn}
\ws{E}_\phi^\mu(\mL)&=\pd{\mL}{(\partial_\mu\phi)}=-\frac{1}{2}\phi^\mu\\
\ws{E}_{\phi^\sigma}^\mu(\mL)&=\pd{\mL}{(\partial_\mu\phi^\sigma)}=\frac{1}{2}\phi\delta^\mu_\sigma.
\end{eqn}

We thus solve for the 4-tuples given by Eq.~(\ref{defFourTuple}):
\begin{eqn}
A^\mu_{P_\alpha}&=\frac{1}{2}\left(\phi^\mu\partial_\alpha\phi-\phi\partial_\alpha\phi^\mu\right)\\
A^\mu_{M_{\alpha\beta}}&=\frac{1}{2}\Bigg(x_\alpha\pd{\phi}{x^\beta}\phi^\mu-x_\beta\pd{\phi}{x^\alpha}\phi^\mu+\phi_\alpha\phi\delta^\mu_\beta-\phi_\beta\phi\delta^\mu_\alpha\\
&\hspace{40pt}-x_\alpha\pd{\phi^\mu}{x^\beta}\phi+x_\beta\pd{\phi^\mu}{x^\alpha}\phi\Bigg).
\label{Tuplesfor4VectorTheory}
\end{eqn}

For each respective symmetry, we again substitute $A$ from Eq.~(\ref{Tuplesfor4VectorTheory}) and ${B=-\mL\zeta}$ from Eq.~(\ref{resultOfVQ2}) into Eq.~(\ref{conciseConsLawForVariational}), to derive 10 conservation laws:
\begin{eqn}
0&=\ws{D}_\mu\Bigg[\frac{1}{2}\Bigg(\phi^\mu\partial^\alpha\phi-\phi\partial^\alpha\phi^\mu\Bigg)+\eta^{\mu\alpha}\mL\Bigg]\\
&\eqdef\ws{D}_\mu T^{\mu\alpha}\\
0&=\ws{D}_\mu\Bigg[x^\alpha T^{\mu\beta}-x^\beta T^{\mu\alpha}+\frac{1}{2}\phi\bigg(\phi^\alpha\eta^{\mu\beta}-\phi^\beta\eta^{\mu\alpha}\bigg)\Bigg]\\
&\eqdef\ws{D}_\mu L^{\mu\alpha\beta}.
\label{consLawFlatFiveVectorNoVierbeinTheory}
\end{eqn}

It is straightforward to show that these conservation laws are equivalent to the conservation laws of Eq.~(\ref{twoConsLawsOfOriginalTheory})---in that they differ by trivial conservation laws, as defined in \cite{olver_textbook_1993}.  (See footnote \ref{footnoteEquivTrivial}.)  In particular, the familiar energy-momenta of scalar theory in Eq.~(\ref{twoConsLawsOfOriginalTheory}) are equivalent to the above energy-momenta of scalar + 4-vector theory.

As such, an experiment measuring a theory's dynamics and conserved currents would not be able to distinguish between a scalar theory and a scalar + 4-vector theory.

\section{Real 5-Vector Field Theory}
In the two foregoing scalar and scalar + 4-vector theories, we have defined the Poincar\'e symmetries to act horizontally---that is, on the spacetime manifold.  We note that, by definition, the target of a Poincar\'e Lie group action must take values in a continuum.  These two observations imply that the prior theories' symmetries are inconsistent with a discretization of spacetime.  This is the fundamental reason that lattice field theories, which define particles at fixed, discrete points embedded in spacetime, fail to preserve Poincar\'e symmetries and their associated invariants.

Although we defer to our companion paper the presentation of a \emph{discrete, Poincar\'e-invariant theory}, we briefly motivate the following section by considering the prerequisites for such a theory.  In light of the preceding argument, the Poincar\'e symmetries of a theory with a discrete horizontal subspace $X$ must act only on the theory's vertical subspace $U^{(n)}$---that is, only on its dependent variables.

One might hope that the vertical evolutionary representatives---${P_\alpha}_Q$ and ${M_{\alpha\beta}}_Q$---of the previous sections would be sufficient for this purpose, but these merely camouflage their horizontal action by including derivatives of the dependent variables in their coefficients.  Indeed, there is an effective sense in which any vector field with a term of the form $f[u]u_x\partial_u$ has an action on the horizontal subspace $X\subset\ws{Jet}(n)$.  After all, $f[u]u_x\partial_u$ is equivalent---in the definition of \cite{olver_textbook_1993}---to the horizontal vector field $f[u]\partial_x$.

This equivalence is more than mere formality.  If we examine ${P_\alpha}_Q=-(\partial\phi/\partial x^\alpha)\partial_\phi$ from Eq.~(\ref{verticalSymms}), for example, we see that the Poincar\'e index $\alpha$ adorns the spacetime coordinate.  A discretization of this coordinate necessarily breaks, therefore, the infinitesimal Poincar\'e invariance of the theory; indeed, any theory set against a background of Poincar\'e-transformable spacetime presupposes a continuous universe.

In this section, therefore, we evolve our theory to discover the `Poincar\'e lift' that properly verticalizes our symmetries.  We proceed in two pedagogical steps:
\begin{enumerate}[label=(\roman*)]
\item We first cleave the Poincar\'e group action from the background coordinates of our theory by introducing a vierbein solder field into the scalar + 4-vector Lagrangian.  In this `vierbein formalism', we find vertical Poincar\'e symmetries of our EOM, but discover that the theory---i.e., the Lagrangian---is not entirely Poincar\'e-symmetric.
\item We then repair this vierbein theory with an edifying matrix formalism we call \emph{5-vector theory}, introducing (a) the 5-vector particle; (b) its antiparticle; and (c) the f\"unfbein solder field.  We demonstrate the vertical Poincar\'e invariance of the 5-vector EOM \emph{and} Lagrangian, and derive the theory's conservation laws.
\end{enumerate}

\subsection{Poincar\'e Lift \#1: Vierbein Formalism}
We begin by taking inspiration from Eq.~(\ref{ScalarPlus4VectorLagrangian}) and define the following Lagrangian:
\begin{eqn}
\mL\defeq\frac{1}{2}\phi^\mu e_\mu^{~a}\eta_{ab}e_\nu^{~b}\phi^\nu-\frac{1}{2}\phi^\mu e_\mu^{~a}\partial_a\phi+\frac{1}{2}\phi e_\nu^{~b}\partial_b\phi^\nu-V(\phi).
\label{FiveVectorLagrangian}
\end{eqn}
In this definition, we have abandoned the Poincar\'e-deformable coordinate system $\{x^\mu\}$ and denote by $\{x^a\}$ the flat background of our theory---a 4-D Cartesian manifold with Minkowski metric $\eta_{ab}$ of signature $($$-$$+$$+$$+$$)$.  (See footnote \ref{whyCartesian}.)  We have facilitated the introduction of these Cartesian coordinates using the vierbein, substituting
\begin{eqn}
\partial_\mu~\rightarrow~e_\mu^{~a}\partial_a
\end{eqn}
in our Lagrangian and thereby cleaving the target indices $\{\mu\}$ of Poincar\'e symmetries from the background coordinate indices $\{a\}$.  We shall often denote a point in this flat background as $x\in\{x^a\}$.

Latin (Cartesian) indices may be lowered with $\eta_{ab}$ and raised with its inverse, such that
\begin{eqn}
\partial^a=\eta^{ab}\partial_b.
\end{eqn}
The repetition of Latin indices indicates a sum---as in the familiar flat-spacetime Einstein summation convention.

Our Lagrangian again has five dynamical variables, in the form of a Lorentz 4-vector $\phi^\mu$ and a scalar field $\phi$.  It additionally has a $4\times4$ matrix vierbein solder field $e_\mu^{~a}$, with inverse $e^\nu_{~b}$:
\begin{eqn}
e_\mu^{~a}e^\mu_{~b}=\delta^a_b~~~\text{and}~~~e_\mu^{~a}e^\nu_{~a}=\delta_\mu^\nu.
\label{vierbeinIdentities}
\end{eqn}
We note for convenience that $\partial_{\left(e^\alpha_{~a}\right)}e_\beta^{~b}=-e_\beta^{~a}e_\alpha^{~b}$, an identity readily derived by differentiating Eq.~(\ref{vierbeinIdentities}).

These fields are defined to be functions over $\{x^a\}$: $\phi(x),$ $\phi^\mu(x)$, $e_\nu^{~b}(x)$, and so on.  Greek (Poincar\'e) indices may be raised and lowered by a `metric' $g_{\mu\nu}$ formed of the vierbeins:
\begin{eqn}
\phi_\mu=g_{\mu\nu}\phi^\nu\equiv e_\mu^{~a}\eta_{ab}e_\nu^{~b}\phi^\nu\\
\phi^\mu=g^{\mu\nu}\phi_\nu\equiv e^\mu_{~a}\eta^{ab}e^\nu_{~b}\phi_\nu.
\label{raiseLower}
\end{eqn}
One may profitably regard $g_{\mu\nu}$ and $g^{\mu\nu}$ simply as compact notations for these arrangements of vierbein fields.

Crucially, in this section we make the following assumptions for the `ungauged' vierbein solder field:
\begin{enumerate}[label=(\alph{enumi})]
\item $e_\mu^{~a}(x)$ is constant---that is, ${\partial_b(e_\mu^{~a})=0}$ $\forall$ $x$;
\item $e_\mu^{~a}$ transforms under global Poincar\'e transformations, in a manner to be defined; and
\item $e_\mu^{~a}$ is a static, non-dynamical field.
\end{enumerate}
These assumptions will be relaxed in the ensuing matrix formalism.  In the present vierbein formalism, however, these \emph{ungauged assumptions} will be applied steadfastly.

We may apply Euler operators---defined in terms of $\{x^a\}$ coordinates---to Eq.~(\ref{FiveVectorLagrangian}) and derive the EOM of our vierbein theory:
\begin{eqn}
(\star):~~0&=\ws{E}_{\phi}(\mL)=-V'(\phi)+e_\mu^{~a}\partial_a\phi^\mu\\
(\star\star)_\sigma:~~0&=\ws{E}_{\phi^\sigma}(\mL)=g_{\sigma\mu}\phi^\mu-e_\sigma^{~a}\partial_a\phi.
\label{theEOM}
\end{eqn}
We find it convenient to refer to these EOM in the analysis below as $(\star)$ and $(\star\star)_\sigma$.  It is worth promptly noting their similarity to the EOM of Eq. (\ref{RealPlusFourVectorEOM}).

We now take the decisive step and lift our Poincar\'e generators to vertical vector fields.  We define them as follows:
\begin{eqn}
P^\alpha&\defeq\phi^\alpha\partial_\phi+\left(\partial^a\right)\partial_{e_\alpha^{~a}}\\
M^{\alpha\beta}&\defeq\phi^\sigma\left(\eta^{\alpha\nu}\delta_\sigma^\beta-\eta^{\beta\nu}\delta_\sigma^\alpha\right)\partial_{\phi^\nu}\\
&\hspace{20pt}-e_\sigma^{~a}\left(\eta^{\alpha\sigma}\delta_\nu^\beta-\eta^{\beta\sigma}\delta_\nu^\alpha\right)\partial_{e_\nu^{~a}}.
\label{VierbeinFormalismGenerators}
\end{eqn}
These differential operators are defined to represent the action of the Poincar\'e generators on our matter and solder fields; we will soon verify that they satisfy the appropriate Lie algebra.

The Poincar\'e vector fields of Eq. (\ref{VierbeinFormalismGenerators}) warrant some examination.  To clarify the \emph{in situ} action of $P^\alpha$, we first note that
\begin{eqn}
P^\alpha[f(\phi^\mu)e_\sigma^{~a}g(\phi^\nu)]=f(\phi^\mu)\delta_\sigma^\alpha\partial^ag(\phi^\nu).
\end{eqn}
The logic for this unusual differential operator $\partial^a$ in the coefficient of the $P^\alpha$ symmetry will be clarified when we replace $P^\alpha$ and $M^{\alpha\beta}$ with improved matrix operators in the forthcoming matrix formalism.

We further note that, unlike the prolonged vertical symmetries defined in Eqs.~(\ref{verticalSymms}) and (\ref{prolongedVertFiveVectorSymmetries}), which are in fact equivalent to horizontal symmetries, we have defined vertical generators in Eq.~(\ref{VierbeinFormalismGenerators}) that do not involve derivatives of dependent variables in their coefficients.

The prolongation of our global symmetries to sufficient order takes the following form:
\begin{eqn}
\ws{pr}[P^\alpha]&=\phi^\alpha\partial_\phi+\left(\partial_a\phi^\alpha\right)\partial_{\left(\partial_a\phi\right)}+\left(\partial^a\right)\partial_{e_\alpha^{~a}}\\
\ws{pr}[M^{\alpha\beta}]&=\phi^\sigma\left(\eta^{\alpha\nu}\delta_\sigma^\beta-\eta^{\beta\nu}\delta_\sigma^\alpha\right)\partial_{\phi^\nu}\\
&\hspace{20pt}+\left(\partial_a\phi^\sigma\right)\left(\eta^{\alpha\nu}\delta_\sigma^\beta-\eta^{\beta\nu}\delta_\sigma^\alpha\right)\partial_{(\partial_a\phi^\nu)}\\
&\hspace{20pt}-e_\sigma^{~a}\left(\eta^{\alpha\sigma}\delta_\nu^\beta-\eta^{\beta\sigma}\delta_\nu^\alpha\right)\partial_{e_\nu^{~a}}.
\label{prolongedVierbeinFormalismGenerators}
\end{eqn}

We may demonstrate that these are infinitesimal symmetries of the Euler-Lagrange equations, as follows:
\begin{eqn}
\ws{pr}[P^\alpha](\star)&=-V''(\phi)g^{\alpha\mu}(\star\star)_\mu+e^\alpha_{~a}\ws{D}^a(\star)\\
&\hspace{17pt}+g^{\alpha\mu}\ws{D}_a\ws{D}^a(\star\star)_\mu-e^\alpha_{~a}e^\mu_{~b}\ws{D}^a\ws{D}^b(\star\star)_\mu\\
\ws{pr}[M^{\alpha\beta}](\star)&=0\\
\ws{pr}[P^\alpha](\star\star)_\sigma&=\delta_\sigma^\alpha e^\mu_{~a}\ws{D}^a(\star\star)_\mu\\
\ws{pr}[M^{\alpha\beta}](\star\star)_\sigma&=[\delta^\alpha_\sigma\eta^{\beta\mu}-\delta^\beta_\sigma\eta^{\alpha\mu}](\star\star)_\mu.
\label{EOMPoincareInvariance}
\end{eqn}

Each right hand side above clearly vanishes on shell---that is, on the submanifold of the jet space satisfying the EOM of Eq.~(\ref{theEOM}).  The flow of the dependent variables along these vector fields therefore carries solutions of our EOM into other solutions, leaving the solution submanifold invariant.  By definition, therefore, $P^\alpha$ and $M^{\alpha\beta}$ of Eq.~(\ref{VierbeinFormalismGenerators}) are symmetries of our EOM.

It is furthermore straightforward to check that these symmetries, as defined in Eq.~(\ref{VierbeinFormalismGenerators}), satisfy the Poincar\'e Lie algebra:
\begin{eqn}
\left\llbracket P^\alpha,P^\beta\right\rrbracket&=0\\
\left\llbracket M^{\alpha\beta},P^\mu\right\rrbracket&=\eta^{\alpha\mu}P^\beta-\eta^{\beta\mu}P^\alpha\\
\left\llbracket M^{\alpha\beta},M^{\mu\nu}\right\rrbracket&=\eta^{\alpha\mu}M^{\beta\nu}+\eta^{\beta\nu}M^{\alpha\mu}-\eta^{\alpha\nu}M^{\beta\mu}-\eta^{\beta\mu}M^{\alpha\nu}.
\label{PoincareLieAlgebra}
\end{eqn}

The Poincar\'e invariance of the above theory is therefore apparent in the 10 symmetries of our EOM---defined in Eq.~(\ref{VierbeinFormalismGenerators}) and verified in Eq.~(\ref{EOMPoincareInvariance})---which satisfy the Poincar\'e Lie algebra in Eq.~(\ref{PoincareLieAlgebra}).

And yet, the Poincar\'e invariance of the vierbein formalism is incomplete.  Calculating ${\ws{pr}[P^\alpha](\mL)}$ and ${\ws{pr}[M^{\alpha\beta}](\mL)}$ for $\mL$ in Eq.~(\ref{FiveVectorLagrangian}), it is easily shown that while $M^{\alpha\beta}$ of Eq.~(\ref{VierbeinFormalismGenerators}) is a variational symmetry of $\mL$---as defined in Eq.~(\ref{defineVarSymm})---$P^\alpha$ is not.  While variational symmetries of a Lagrangian $\mL$ are always symmetries of its EOM ${\ws{E}_u(\mL)}$, the converse is not always true \cite{olver_textbook_1993}, as we have just seen.

In the subsequent 5-vector theory, we will find a completely Poincar\'e invariant theory by addressing the following three limitations of our vierbein formalism:
\begin{itemize}
\item The vierbein's transformation under $P^\alpha$ of Eq.~(\ref{VierbeinFormalismGenerators}) is too inflexible, in a sense that will be clarified.  We will use the matrix formalism to define a more flexible notion of our solder field's translation, which will require the vierbein's expansion into the $5\times5$ f\"unfbein.
\item The vierbein theory lacks an antiparticle.  If we consider the EOM of Eq.~(\ref{theEOM}), we note that the momentum of our $\phi$ field is---roughly speaking---characterized by the $\phi^\mu$ field.  Because the 5-vector must transform under translations, its `linear momentum charge'---in a quantized theory---should be conserved in its interactions.  The terms of a 5-vector Lagrangian should therefore couple only the 5-vector field and its antiparticle.
\item The vierbein field is non-dynamical.  The setting of Noether's procedure in jet space requires that any vertical field---any dependent variable---has a corresponding Euler-Lagrange equation.  We shall therefore define our f\"unfbein solder field to be dynamical.
\end{itemize}

In the following matrix formalism, we render the improvements motivated above.  We will introduce three new elements into our Lagrangian: (a) the \emph{5-vector} particle; (b) its antiparticle---the \emph{twisted 5-vector}; and (c) the \emph{f\"unfbein} solder field.

\subsection{Poincar\'e Lift \#2: Matrix Formalism}

We therefore refine our vierbein theory by reexpressing its fields and group actions in the form of matrix representations. In the following, we recast the action of the Poincar\'e symmetries---represented as differential operators on jet space in Eq.~(\ref{VierbeinFormalismGenerators})---as matrix transformations on our matter and solder fields.

We choose the following faithful, non-unitary $5\times5$ matrix representation $\rho$ of the Poincar\'e group:
\begin{eqn}
\rho:\left\{
\begin{alignedat}{3}
&(\Lambda,\vphi)&&\rightarrow~~\gv{\Lambda}&&\equiv\left[\begin{matrix}\Lambda^\mu_{~\nu}&\v{0}\\\vphi_\nu&1\end{matrix}\right]\\
&(\mOne,\v{0})&&\rightarrow~~\gv{\Lambda}_0&&\equiv\left[\begin{matrix}\delta^\mu_\nu&\v{0}\\\v{0}&1\end{matrix}\right]\\
&(\Lambda,\vphi)^{-1}&&\rightarrow~~\gv{\Lambda}^{-1}&&\equiv\left[\begin{matrix}(\Lambda^{-1})^\mu_{~\nu}&\v{0}\\-\vphi_\sigma(\Lambda^{-1})^\sigma_{~\nu}&1\end{matrix}\right].
\end{alignedat}\right.
\end{eqn}
We correspondingly organize our matter fields into a 5-component column vector that we call a \emph{5-vector}---$\gv{\phi}$---whose components and Poincar\'e transformation are defined as follows:
\begin{eqn}
\gv{\phi}(x)&\defeq\left[\begin{matrix}[l]\phi^\mu\\~\phi\end{matrix}\right](x)\\
\gv{\Lambda}\cdot\gv{\phi}&\defeq\left[\begin{matrix}\Lambda^\mu_{~\nu}&\v{0}\\\vphi_\nu&1\end{matrix}\right]\cdot\left[\begin{matrix}\phi^\nu\\\phi\end{matrix}\right]=\left[\begin{matrix}\Lambda^\mu_{~\nu}\phi^\nu\\\phi+\vphi_\nu\phi^\nu\end{matrix}\right].
\label{5VectorDefn}
\end{eqn}

We furthermore augment our vierbein into a $5\times5$ matrix we call the \emph{f\"unfbein}---$\v{e}$---whose components and transformation are defined to be:
\begin{eqn}
\v{e}(x)\defeq&\left[\begin{matrix}e_\mu^{~a}&\v{0}\\e_\mu&1\end{matrix}\right](x)\\
\v{e}\cdot\gv{\Lambda}^{-1}\defeq&\left[\begin{matrix}e_\mu^{~a}&\v{0}\\e_\mu&1\end{matrix}\right]\cdot\left[\begin{matrix}\Lambda^\mu_{~\nu}&\v{0}\\\vphi_\nu&1\end{matrix}\right]^{-1}\\
=&\left[~\begin{matrix}[c|c]e_\mu^{~a}(\Lambda^{-1})^\mu_{~\nu}&\v{0}\\\hline e_\mu(\Lambda^{-1})^\mu_{~\nu}-\vphi_\sigma(\Lambda^{-1})^\sigma_{~\nu}&1\end{matrix}~\right].
\label{FunfbeinDefn}
\end{eqn}
We call particular attention to the \emph{inverse} action of $\gv{\Lambda}$ on the f\"unfbein, as well as the distinct use of left and right matrix multiplication for the 5-vector and f\"unfbein, respectively.  For completeness, we note that $\v{e}^T$ transforms just as it should: ${\v{e}^T\rightarrow\left(\v{e}\cdot\gv{\Lambda}^{-1}\right)^T=\gv{\Lambda}^{-T}\v{e}^T}$.

We define our antiparticle---the \emph{twisted 5-vector} ${\gv{\tphi}}$---as a 5-component row vector with the following Poincar\'e transformation:
\begin{eqn}
\gv{\tphi}(x)&\defeq\left[\tphi^\mu~\tphi\right](x)\\
\gv{\tphi}\cdot\gv{\Lambda}^T&\defeq\left[\tphi^\nu~\tphi\right]\cdot\left[\begin{matrix}\Lambda^\mu_{~\nu}&\vphi_\nu\\\v{0}&1\end{matrix}\right]=\left[\tphi^\nu\Lambda^\mu_{~\nu}~\Big|~\tphi+\tphi^\nu\vphi_\nu\right].
\label{A5VectorDefn}
\end{eqn}
Despite the identical Poincar\'e transformations of $\gv{\tphi}$ and $\gv{\phi}^T$, the EOM of $\gv{\tphi}$ will emphasize its uniqueness from $\gv{\phi}$, as we shall see.

As a point of clarification, our notation in Eq.~(\ref{A5VectorDefn}) is somewhat schematic.  One might prefer to explicitly notate the implied transpose of the 4-vector $\tphi^\mu$ and the Poincar\'e matrix components, e.g. $[(\tphi^\nu)^T(\Lambda^\mu_{~\nu})^T]$.  Wherever components of $\gv{\tphi}$, $\gv{\Lambda}^T$ and $\v{e}^T$ are written, however, we will continue to rely on indices to indicate the appropriate order of operations, as we have in Eq.~(\ref{A5VectorDefn}), and we trust that it will not be a source of confusion.

We observe that, given the matrix group actions defined in Eqs.~(\ref{5VectorDefn})-(\ref{A5VectorDefn}), the expressions
\begin{eqn}
\gv{\tphi}\cdot\v{e}^T=\left[\tphi^\mu~\tphi\right]\cdot\left[\begin{matrix}e_\mu^{~a}&e_\mu\\\v{0}&1\end{matrix}\right]~~~\text{and}~~~\v{e}\cdot\gv{\phi}=\left[\begin{matrix}e_\mu^{~a}&\v{0}\\e_\mu&1\end{matrix}\right]\cdot\left[\begin{matrix}\phi^\mu\\\phi\end{matrix}\right]
\label{invarPoincareExpressions}
\end{eqn}
are invariant under Poincar\'e transformations.

As a final introductory note before we define ungauged 5-vector theory, we state our \emph{solder field assumptions} for the `ungauged' f\"unfbein in the matrix formalism:
\begin{enumerate}[label=(\alph{enumi})]
\item $\v{e}(x)$ is constant\footnote{This assumption is applied as if it were an \emph{on-shell condition}.  It is not required a priori, for example, when deriving EOM or applying Poincar\'e transformations.  In general relativity, a related assumption appears as the `tetrad postulate' \cite{carroll_textbook_2003}, wherein ${\nabla_a(e_\mu^{~b})=0}$ for the covariant derivative $\nabla_a$.  Since our present theory is as yet ungauged, it is analogous that we should require ${\partial_a\v{e}=0}$.}---that is, ${\partial_a\v{e}=0}$ $\forall$ $x$;
\item $\v{e}$ transforms under global Poincar\'e transformations, as defined in Eq.~(\ref{FunfbeinDefn}); and
\item $\v{e}$ is a dynamical field, (though its EOM will turn out to be indeterminate in this as-yet-ungauged theory).
\end{enumerate}
In (c), we have relaxed the non-dynamical solder field assumption of the vierbein formalism.

We are at last ready to define the \emph{5-vector Lagrangian} in our matrix formalism:
\begin{eqn}
\mL&\defeq\gv{\tphi}\v{e}^T\Big\{\gv{\partial}_\eta-\gv{\partial}_{\text{\tiny{$V$}}}\Big\}\v{e}\gv{\phi}\\
&\defeq\left[\tphi^\mu~\tphi\right]\left[\begin{matrix}e_\mu^{~a}&e_\mu\\\v{0}&1\end{matrix}\right]\cdot\\
&\hspace{30pt}\left\{\left[\begin{matrix}\eta_{ab}&-\partial_a\\-\partial_b&\partial^2\end{matrix}\right]-\left[\begin{matrix}\mZero&\v{0}\\\v{0}&\partial^2-m^2\end{matrix}\right]\right\}\left[\begin{matrix}e_\nu^{~b}&\v{0}\\e_\nu&1\end{matrix}\right]\left[\begin{matrix}\phi^\nu\\\phi\end{matrix}\right].
\label{theLagrangian}
\end{eqn}
Defining the Poincar\'e-invariant quantities
\begin{eqn}
\phi_0&\defeq\phi+e_\mu\phi^\mu\\
\tphi_0&\defeq\tphi+e_\mu\tphi^\mu,
\label{PoincareInvariantQtys}
\end{eqn}
we may concisely rewrite this Lagrangian as:
\begin{eqn}
\mL=\tphi^\mu g_{\mu\nu}\phi^\nu-\tphi^\mu e_\mu^{~a}\partial_a\phi_0-\tphi_0\partial_b(e_\nu^{~b}\phi^\nu)+m^2\tphi_0\phi_0.
\label{5VectorUngaugedLagrangian}
\end{eqn}
We now make several comments about $\mL$.

First, we have restricted our theory to the (almost Klein-Gordon) potential ${V(\tphi_0,\phi_0)=m^2\tphi_0\phi_0}$.  This potential is uniquely well-suited to our bilinear matrix formulation, however, self-interacting potentials ${V=V(\tphi_0,\phi_0)}$, defined in terms of the Poincar\'e-invariant quantities of Eq.~(\ref{PoincareInvariantQtys}), are equally admissible.  We also note that, given the positive sign of $V$ within $\mL$, we have arbitrarily chosen to flip the overall sign of the Lagrangian.

Second, we have included in Eq.~(\ref{theLagrangian}) two matrices formed of `background $\{x^a\}$ operators', denoted by the symbols $\gv{\partial}_\eta$ and $\gv{\partial}_{\text{\tiny{$V$}}}$.  Importantly, these matrices do not transform under Poincar\'e transformation.  Loosely speaking, they characterize the \emph{horizontal kinematics} of our fields on $\{x^a\}$, while the other matrices of our Lagrangian characterize the \emph{vertical dynamics} of our fields.

There is a redundancy in the formulation of these matrices: Rather than combine $\gv{\partial}_\eta$ and $\gv{\partial}_{\text{\tiny{$V$}}}$ into a single matrix, we include second derivatives ${\partial^2\equiv\partial^a\partial_a}$ in $\gv{\partial}_\eta$ and $\gv{\partial}_{\text{\tiny{$V$}}}$ whose terms ${\tphi_0\partial^2\phi_0}$ cancel each other out.

We do this to highlight the relationship between the Poincar\'e transformation of the f\"unfbein, and the jet space transformations given in the vierbein formalism's Eq.~(\ref{VierbeinFormalismGenerators}).  Taking ${\gv{\partial}_\eta}$ alone, the Poincar\'e translation ${[\v{e}^T\gv{\partial}_\eta\v{e}]\rightarrow[\gv{\Lambda}^{-T}\v{e}^T\gv{\partial}_\eta\v{e}\gv{\Lambda}^{-1}]}$ for ${\gv{\Lambda}=(\mOne,\vphi)}$ exactly reproduces the vierbein translation $P^\alpha$ previously defined in Eq.~(\ref{VierbeinFormalismGenerators}).  Indeed, it is now clear why the unusual coefficient $\partial^a$ appears in the definition of $P^\alpha$ in the vierbein formalism: It effectively replaces the transformation of the four \emph{hidden} $e_\mu$ components of the f\"unfbein field, without explicitly including them in the theory.

In this sense, the transformation of $\v{e}$ in the full context of ${[\v{e}^T\{\gv{\partial}_\eta-\gv{\partial}_V\}\v{e}]}$ reveals the flexibility of the f\"unfbein's transformations relative to the vierbein.

Having characterized elements of our 5-vector Lagrangian, we now proceed to examine its Poincar\'e symmetry.  The matrix generators of the Poincar\'e group are given by:
\begin{eqn}
[P^\alpha]^\mu_{~\sigma}&\defeq\left[\begin{matrix}\mZero&\v{0}\\\delta^\alpha_\sigma&0\end{matrix}\right]\\
[M^{\alpha\beta}]^\mu_{~\sigma}&\defeq\left[\begin{matrix}\left(\delta^\alpha_\sigma\eta^{\beta\mu}-\delta^\beta_\sigma\eta^{\alpha\mu}\right)&\v{0}\\\v{0}&0\end{matrix}\right].
\label{MatrixFormalismGenerators}
\end{eqn}
It is easily verified that these generators satisfy the Lie algebra of Eq.~(\ref{PoincareLieAlgebra}).

We apply these generators to $\mL$ of Eq.~(\ref{theLagrangian}) as we would differential operators, and note that the prolongations of their actions are equivalent to transforming our matrix fields wherever those fields appear---even under a derivative operator.  In particular, the matrix generators of Eq.~(\ref{MatrixFormalismGenerators})---(and their negations and transposes, as appropriate)---applied to the fields of $\mL$ \emph{in situ}, generate the same transformations as the following prolonged vector fields:
\begin{eqn}
\ws{pr}[P^\alpha]&=\smashoperator{\sum\limits_{J=\{\emptyset,a,\cdots\}}}\Big[\phi^\alpha_J\partial_{\phi_J}+\tphi^\alpha_J\partial_{\tphi_J}\Big]-\partial_{e_\alpha}\\
\ws{pr}[M^{\alpha\beta}]&=\smashoperator{\sum\limits_{J=\{\emptyset,a,\cdots\}}}\Big[\phi^\sigma_J(\delta^\alpha_\sigma\eta^{\beta\nu}-\delta^\beta_\sigma\eta^{\alpha\nu})\partial_{\phi^\nu_J}\\
&\hspace{30pt}+\tphi^\sigma_J(\delta^\alpha_\sigma\eta^{\beta\nu}-\delta^\beta_\sigma\eta^{\alpha\nu})\partial_{\tphi^\nu_J}\\
&\hspace{30pt}-(e_\sigma)_J(\delta^\alpha_\nu\eta^{\beta\sigma}-\delta^\beta_\nu\eta^{\alpha\sigma})\partial_{(e_\nu)_J}\\
&\hspace{30pt}-(e_\sigma^{~c})_J(\delta^\alpha_\nu\eta^{\beta\sigma}-\delta^\beta_\nu\eta^{\alpha\sigma})\partial_{(e_\nu^{~c})_J}\Big].
\end{eqn}

We therefore observe that the 5-vector Lagrangian of Eq.~(\ref{theLagrangian}) is completely invariant under Poincar\'e transformations:
\begin{eqn}
\ws{pr}[P^\alpha](\mL)&=0\\
\ws{pr}[M^{\alpha\beta}](\mL)&=0.
\label{showMatrixVarSymm}
\end{eqn}

Eq.~(\ref{showMatrixVarSymm}) demonstrates that our matrix transformations are indeed variational symmetries of our Lagrangian in Eq.~(\ref{theLagrangian})---and that we have, therefore, repaired the vierbein formalism.  Having successfully lifted the Poincar\'e symmetries of our classical field theory, we have thus fulfilled our `verticalization program'.

We may now derive the EOM for our matter and solder fields.  We simply apply Euler operators---again defined in terms of $\{x^a\}$ coordinates---to the Lagrangian of Eq.~(\ref{theLagrangian}):
\begin{eqn}
\hspace{-5pt}&\v{0}=\ws{E}_{\tiny{\text{$\gv{\tphi}$}}}(\mL)=\v{e}^T\big\{\gv{\partial}_\eta-\gv{\partial}_{\text{\tiny{$V$}}}\big\}\v{e}\gv{\phi}\\
\hspace{-5pt}&\v{0}=\ws{E}_{\tiny{\text{$\gv{\phi}$}}}(\mL)=\v{e}^T\big\{\tilde{\gv{\partial}}_\eta-\gv{\partial}_{\text{\tiny{$V$}}}\big\}\v{e}\gv{\tphi}^T\\
\hspace{-5pt}&0=\ws{E}_{e_\sigma}(\mL)=-\tphi^\sigma\left[\partial_b(e_\nu^{~b}\phi^\nu)-m^2(e_\nu\phi^\nu+\phi)\right]\\
\hspace{-5pt}&\hspace{57pt}+\phi^\sigma\left[\partial_a(e_\mu^{~a}\tphi^\mu)+m^2(e_\mu\tphi^\mu+\tphi)\right]\\
\hspace{-5pt}&0=\ws{E}_{e_\sigma^{~a}}(\mL)=\tphi^\sigma\eta_{ab}\left[e_\nu^{~b}\phi^\nu-\partial^b\left(e_\nu\phi^\nu+\phi\right)\right]\\
\hspace{-5pt}&\hspace{52pt}+\phi^\sigma\eta_{ab}\left[e_\mu^{~b}\tphi^\mu+\partial^b\left(e_\mu\tphi^\mu+\tphi\right)\right].
\label{theUngaugedEOM}
\end{eqn}
In the second equation above, we have employed a new `kinematic matrix':
\begin{eqn}
\tilde{\gv{\partial}}_\eta\equiv\left[\begin{matrix}\eta_{ab}&\partial_a\\\partial_b&\partial^2\end{matrix}\right],
\end{eqn}
which reflects the change in sign of first-order derivatives upon the evaluation of an Euler operator.

We define the following Poincar\'e-invariant quantities
\begin{eqn}
(\circ)&\defeq m^2\phi_0-\partial_b(e_\nu^{~b}\phi^\nu)\\
(\bullet)&\defeq m^2\tphi_0+\partial_b(e_\mu^{~b}\tphi^\mu)\\
(\circ\circ)_a&\defeq\eta_{ab}e_\nu^{~b}\phi^\nu-\partial_a\phi_0\\
(\bullet\bullet)_a&\defeq\eta_{ab}e_\mu^{~b}\tphi^\mu+\partial_a\tphi_0,
\end{eqn}
so that we may reexpress our EOM of Eq.~(\ref{theUngaugedEOM}) as follows:
\begin{eqn}
\hspace{-5pt}&0=\ws{E}_{\tphi^\sigma}(\mL)=e_\sigma^{~a}(\circ\circ)_a+e_\sigma(\circ)\\
\hspace{-5pt}&0=\ws{E}_{\tphi}(\mL)=(\circ)\\
\hspace{-5pt}&0=\ws{E}_{\phi^\sigma}(\mL)=e_\sigma^{~a}(\bullet\bullet)_a+e_\sigma(\bullet)\\
\hspace{-5pt}&0=\ws{E}_{\phi}(\mL)=(\bullet)\\
\hspace{-5pt}&0=\ws{E}_{e_\sigma}(\mL)=\tphi^\sigma(\circ)+\phi^\sigma(\bullet)\\
\hspace{-5pt}&0=\ws{E}_{e_\sigma^{~a}}(\mL)=\tphi^\sigma(\circ\circ)_a+\phi^\sigma(\bullet\bullet)_a.
\label{eulerOperator5VectorEOM}
\end{eqn}
We see that all of our EOM are therefore solved when:
\begin{eqn}
(\circ)=(\bullet)=(\circ\circ)_a=(\bullet\bullet)_a=0.
\label{theBriefUngaugedEOM}
\end{eqn}

We observe that our EOM are indeterminate of the solder field's 20 degrees of freedom (DOF).  This is to be expected in an ungauged theory---inasmuch as the `dynamics' of $\eta_{\mu\nu}$ are `indeterminate' in a flat theory of gravity without curvature.  Indeed, our assumption ${\partial_a\v{e}=0}$ requires $\v{e}$ to be constant over all $\{x^a\}$.

We further note from Eq.~(\ref{theBriefUngaugedEOM}) that \emph{both} $\phi_0$ and $\tphi_0$ obey the Klein-Gordon equation on shell:
\begin{eqn}
(\partial^2-m^2)\phi_0=(\partial^2-m^2)\tphi_0=0.
\label{KGeqn}
\end{eqn}
Despite these identical dynamics, the $(\bullet)$ and $(\bullet\bullet)_a$ EOM for $\gv{\tphi}$ have an important sign difference with respect to $(\circ)$ and $(\circ\circ)_a$.  The \emph{internal} dynamics of the components of $\gv{\tphi}$ are distinctly opposite those of $\gv{\phi}$, encouraging its interpretation as the antiparticle of $\gv{\phi}$.

Finally, we proceed to develop the 10 conservation laws of 5-vector theory.  Because our symmetries are already vertical, and since $B$ of Eq.~(\ref{conciseConsLawForVariational}) exactly vanishes in Eq.~(\ref{showMatrixVarSymm}), we need only calculate the  4-tuple $A$ that completes our conservation laws, as defined in Eq.~(\ref{defFourTuple}).  Gathering the nonzero data required, we find:
\begin{eqn}
\ws{E}_\phi^a(\mL)&=-e_\mu^{~a}\tphi^\mu\\
\ws{E}_{\phi^\sigma}^a(\mL)&=-e_\sigma^{~a}\tphi_0-e_\sigma e_\mu^{~a}\tphi^\mu\\
\ws{E}_{e_\sigma}^{a}(\mL)&=-e_\mu^{~a}\tphi^\mu\phi^\sigma\\
\ws{E}_{e_\sigma^{~b}}^{a}(\mL)&=-\delta^a_b\phi^\sigma\tphi_0\\
Q^{P^\alpha}_\phi&=\phi^\alpha\hspace{28pt}Q^{M^{\alpha\beta}}_{\phi^\nu}=\left(\delta_\sigma^\alpha\eta^{\beta\nu}-\delta_\sigma^\beta\eta^{\alpha\nu}\right)\phi^\sigma\\
Q^{P^\alpha}_{\tphi}&=\tphi^\alpha\hspace{28pt}Q^{M^{\alpha\beta}}_{\tphi^\nu}=\left(\delta_\sigma^\alpha\eta^{\beta\nu}-\delta_\sigma^\beta\eta^{\alpha\nu}\right)\tphi^\sigma\\
Q^{P^\alpha}_{e_\sigma}&=-\delta^\alpha_\sigma\hspace{22pt}Q^{M^{\alpha\beta}}_{e_\nu}=\left(\delta_\nu^\beta\eta^{\alpha\sigma}-\delta_\nu^\alpha\eta^{\beta\sigma}\right)e_\sigma\\
&\hspace{54pt}Q^{M^{\alpha\beta}}_{e_\nu^{~a}}=\left(\delta_\nu^\beta\eta^{\alpha\sigma}-\delta_\nu^\alpha\eta^{\beta\sigma}\right)e_\sigma^{~a}.
\end{eqn}
Following Eq.~(\ref{defFourTuple}), these data yield the following 4-tuples:
\begin{eqn}
A^a_{P^\alpha}&=0\\
A^a_{M^{\alpha\beta}}&=0.
\end{eqn}
We have therefore discovered that the \emph{canonical} Noether currents of 5-vector theory are \emph{trivial}---that is, they vanish on shell.

By Noether's second theorem, this triviality can be seen as the result of the gauge-like Poincar\'e symmetry of $\mL$ in Eq.~(\ref{theLagrangian})---that is, its \emph{local} Poincar\'e symmetry.  Analogous results are found in other locally symmetric (gauge) theories, for example, in the `strong' conservation laws of general relativity \cite{,fletcher_local_1960,goldberg_invariant_1980,kosmann-schwarzbach_noether_2011}.

Nevertheless, it is clear from the comparable physics of scalar + 4-vector theory and 5-vector theory that 5-vector theory also conserves 10 nontrivial currents as it evolves through its EOM solution subspace.  These nontrivial conservation laws may be identified by simply noting that the EOM of Eq.~(\ref{theBriefUngaugedEOM}) have the same structure as the scalar + 4-vector EOM of Eq.~(\ref{RealPlusFourVectorEOM}).

We can therefore write down corresponding conservation laws, taking inspiration from Eq.~(\ref{consLawFlatFiveVectorNoVierbeinTheory}) and recalling that we have sign changes from the overall Lagrangian and antiparticle terms:
\begin{eqn}
\hspace{-15pt}0&=\ws{D}_a\Bigg[e^\alpha_{~b}\bigg(e_\mu^{~a}\tphi^\mu\partial^b\phi_0+\tphi_0\partial^b(e_\mu^{~a}\phi^\mu)\bigg)+e^\alpha_{~b}\eta^{ba}\mL\Bigg]\\
&\eqdef\ws{D}_aT^{a\alpha}\\
\hspace{-15pt}0&=\ws{D}_a\Bigg[\bigg(e^\alpha_{~b}x^bT^{a\beta}-e^\beta_{~b}x^bT^{a\alpha}\bigg)-\eta^{ab}\bigg(\tphi^\alpha e^\beta_{~b}-\tphi^\beta e^\alpha_{~b}\bigg)\phi_0\Bigg]\\
&\eqdef\ws{D}_aL^{a\alpha\beta}.
\label{consLawFiveVectorMatrixTheory}
\end{eqn}
We note that on-shell substitutions from the EOM of Eq.~(\ref{theBriefUngaugedEOM}) reduce the first of these conservation laws to the following equality: ${0=\left(\partial_ae^\alpha_{~b}\right)\cdot\left[\tphi_0\partial^a\partial^b\phi_0-\partial^a\tphi_0\partial^b\phi_0\right]}$.  We therefore use our flat solder field assumption---${\partial_a\v{e}=0}$---to validate that ${\ws{D}_aT^{a\alpha}=0}$.  The second relation, ${\ws{D}_aL^{a\alpha\beta}=0}$, follows similarly.

Before concluding our discussion of ungauged 5-vector theory's conservation laws, we seek to reexpress Eq.~(\ref{consLawFiveVectorMatrixTheory}).  In the manner of a quantum mechanical probability current, we symmetrize the contributions of the 5-vector $\gv{\phi}$ and the twisted 5-vector $\gv{\tphi}$ to the energy-momentum $T^{a\alpha}$.  We further note that the Lagrangian ${\mL=0}$ on shell, so that it may be omitted from $T^{a\alpha}$.  With these considerations, we discover the following modified conservation laws:
\begin{eqn}
0&=\ws{D}_a\bar{T}^{a\alpha}\defeq\ws{D}_a\Bigg[e^\alpha_{~b}e_\mu^{~a}\bigg(\tphi^\mu\partial^b\phi_0-\phi^\mu\partial^b\tphi_0\bigg)\\
&\hspace{74pt}+e^\alpha_{~b}\bigg(\tphi_0\partial^b(e_\mu^{~a}\phi^\mu)-\phi_0\partial^b(e_\mu^{~a}\tphi^\mu)\bigg)\Bigg]\\
0&=\ws{D}_a\bar{L}^{a\alpha\beta}\defeq\ws{D}_a\Bigg[e^\alpha_{~b}x^b\bar{T}^{a\beta}-e^\beta_{~b}x^b\bar{T}^{a\alpha}\Bigg].
\label{finalConsLawFiveVectorMatrixTheory}
\end{eqn}
These symmetrized energy-momenta, which are readily verified by substitutions from Eq.~(\ref{theBriefUngaugedEOM}), will prove indispensable in our discrete companion paper.  We further observe that the symmetrized angular momentum conservation law no longer requires an additional offsetting term, as appeared in Eqs.~(\ref{consLawFlatFiveVectorNoVierbeinTheory}) and (\ref{consLawFiveVectorMatrixTheory}).

\subsection{The Comparable Symmetries of\\5-Vector Theory and Scalar Theory}
We have demonstrated the lifted Poincar\'e symmetries of 5-vector theory, and discovered its nontrivial conservation laws.  We now pause to compare the vertical Poincar\'e symmetries we have defined for 5-vector theory with the more familiar Poincar\'e transformations of scalar field theory.

In Eq.~(\ref{5VectorDefn}), the action of the translation symmetry $P^\alpha$ on our 5-vector particle is reminiscent of the familiar translation of scalar theory---truncated to first order:
\begin{align*}
&\text{Scalar:}&&\phi&&\rightarrow~~\phi+\vphi^\mu\partial_\mu\phi+\cdots\\
&\text{5-Vector:}&&[\phi_\mu~\phi]&&\rightarrow~~\left[\phi_\mu~\Big|~\phi+\vphi^\mu\phi_\mu\right].
\end{align*}

Similarly, inasmuch as the vierbein is a map from the Cartesian background to the local frame---${\partial^\mu=e^\mu_{~a}\partial^a}$---we may compare its transformation under $P^\alpha$ in the differential operator formalism of Eq.~(\ref{VierbeinFormalismGenerators}) with the more familiar transformation of the spacetime derivative under a translation in scalar field theory:
\begin{align*}
&\text{Spacetime Derivative:}&&\partial^\mu&&\rightarrow~~\partial^\mu+\vphi^\nu\partial_\nu\partial^\mu+\cdots\\
&\text{Vierbein:}&&e^\mu_{~a}\partial^a&&\rightarrow~~e^\mu_{~a}\partial^a+\vphi^\mu\partial_a\partial^a.
\end{align*}
Up to indexation, this translation resembles the first-order truncation of its scalar theory counterpart.

The Lorentz symmetry $M^{\alpha\beta}$ in Eq.~(\ref{5VectorDefn}) is even more immediately recognizable from scalar theory than $P^\alpha$.  Indeed, $M^{\alpha\beta}$ effects the transformation of our 4-vectors---that is, the `Lorentz sectors' of our 5-component fields---just as it does for the 4-vector spacetime derivatives of scalar field theory.

\section{Conclusion}

In 5-vector theory, we have thus discovered a Poincar\'e-symmetric theory with vertical group transformations, as desired.  In Eq.~(\ref{theLagrangian}), we have defined the physics of 5-vector theory, and found that its dynamics on a static Cartesian background replicate those of a scalar field in flat spacetime, as in Eq.~(\ref{KGeqn}).  We have furthermore discovered its conservation laws, as expressed in Eq.~(\ref{finalConsLawFiveVectorMatrixTheory}).  We conclude that 5-vector theory is a viable classical field theory, whose dynamics and conservation laws essentially reproduce the physics of a real scalar field.

Upon reflection, a revision of our physical intuition is prompted by this theory.  We have demonstrated how the `background canvas' of 5-vector theory can be regarded as an invariant object---without sacrificing any consequential aspect of the spacetime symmetries of scalar theory.  By constructing the 5-vector and solder field to have Lorentz and translation components, we have recast `background' spacetime symmetries as `foreground' symmetries of dynamical fields.

We furthermore observe that the invariant background $\{x^a\}$ constitutes an absolute reference frame, and appears to restore the notion of simultaneity to relativistic field theory.  After all, two events $x,y\in\{x^a\}$ that satisfy ${x^0=y^0}$ are, formally, simultaneous.

However, a `5-vector observer'---that is, an observer composed of 5-vector matter fields---whose relativistic dynamics are described by Eq.~(\ref{theBriefUngaugedEOM}), would experience the passage of time in her own reference frame.  Therefore, while two events may have a well-defined simultaneity in the Cartesian background, they may not be observed to be simultaneous by such a `vertical' observer.

To lift the Poincar\'e symmetries, 5-vector theory requires the coexistence of the `absolute universal clock' of Newtonian physics, and the `local relativistic clock' of Einsteinian physics.  In our companion paper, we will further demonstrate that this absolute Newtonian clock might well be digital.

\section{Acknowledgments}
This research was supported by the U.S. Department of Energy (DE-AC02-09CH11466).

%


\begin{thebibliography}{8}%
\makeatletter
\providecommand \@ifxundefined [1]{%
 \@ifx{#1\undefined}
}%
\providecommand \@ifnum [1]{%
 \ifnum #1\expandafter \@firstoftwo
 \else \expandafter \@secondoftwo
 \fi
}%
\providecommand \@ifx [1]{%
 \ifx #1\expandafter \@firstoftwo
 \else \expandafter \@secondoftwo
 \fi
}%
\providecommand \natexlab [1]{#1}%
\providecommand \enquote  [1]{``#1''}%
\providecommand \bibnamefont  [1]{#1}%
\providecommand \bibfnamefont [1]{#1}%
\providecommand \citenamefont [1]{#1}%
\providecommand \href@noop [0]{\@secondoftwo}%
\providecommand \href [0]{\begingroup \@sanitize@url \@href}%
\providecommand \@href[1]{\@@startlink{#1}\@@href}%
\providecommand \@@href[1]{\endgroup#1\@@endlink}%
\providecommand \@sanitize@url [0]{\catcode `\\12\catcode `\$12\catcode
  `\&12\catcode `\#12\catcode `\^12\catcode `\_12\catcode `\%12\relax}%
\providecommand \@@startlink[1]{}%
\providecommand \@@endlink[0]{}%
\providecommand \url  [0]{\begingroup\@sanitize@url \@url }%
\providecommand \@url [1]{\endgroup\@href {#1}{\urlprefix }}%
\providecommand \urlprefix  [0]{URL }%
\providecommand \Eprint [0]{\href }%
\providecommand \doibase [0]{http://dx.doi.org/}%
\providecommand \selectlanguage [0]{\@gobble}%
\providecommand \bibinfo  [0]{\@secondoftwo}%
\providecommand \bibfield  [0]{\@secondoftwo}%
\providecommand \translation [1]{[#1]}%
\providecommand \BibitemOpen [0]{}%
\providecommand \bibitemStop [0]{}%
\providecommand \bibitemNoStop [0]{.\EOS\space}%
\providecommand \EOS [0]{\spacefactor3000\relax}%
\providecommand \BibitemShut  [1]{\csname bibitem#1\endcsname}%
\let\auto@bib@innerbib\@empty
\bibitem [{\citenamefont {Glasser}\ and\ \citenamefont
  {Qin}(2019)}]{glasser_discrete5vectortheory}%
  \BibitemOpen
  \bibfield  {author} {\bibinfo {author} {\bibfnamefont {A.~S.}\ \bibnamefont
  {Glasser}}\ and\ \bibinfo {author} {\bibfnamefont {H.}~\bibnamefont {Qin}},\
  }\href@noop {} {\emph {\bibinfo {title} {Restoring Poincar\'e Symmetry to the
  Lattice}}}\ (\bibinfo {year} {in preparation, 2019})\BibitemShut {NoStop}%
\bibitem [{\citenamefont {Wise}(2010)}]{wise_macdowell-mansouri_2010}%
  \BibitemOpen
  \bibfield  {author} {\bibinfo {author} {\bibfnamefont {D.~K.}\ \bibnamefont
  {Wise}},\ }\href {\doibase 10.1088/0264-9381/27/15/155010} {\bibfield
  {journal} {\bibinfo  {journal} {Classical and Quantum Gravity}\ }\textbf
  {\bibinfo {volume} {27}},\ \bibinfo {pages} {155010} (\bibinfo {year}
  {2010})},\ \bibinfo {note} {arXiv: gr-qc/0611154}\BibitemShut {NoStop}%
\bibitem [{\citenamefont {Olver}(1993)}]{olver_textbook_1993}%
  \BibitemOpen
  \bibfield  {author} {\bibinfo {author} {\bibfnamefont {P.~J.}\ \bibnamefont
  {Olver}},\ }\href@noop {} {\emph {\bibinfo {title} {Applications of Lie
  Groups to Differential Equations}}},\ \bibinfo {edition} {2nd}\ ed.,\
  \bibinfo {series} {Graduate Texts in Mathematics}, Vol.\ \bibinfo {volume}
  {107}\ (\bibinfo  {publisher} {Springer-Verlag New York},\ \bibinfo {year}
  {1993})\BibitemShut {NoStop}%
\bibitem [{\citenamefont {Kibble}(1967)}]{kibble_symmetry_1967}%
  \BibitemOpen
  \bibfield  {author} {\bibinfo {author} {\bibfnamefont {T.~W.~B.}\
  \bibnamefont {Kibble}},\ }\href {\doibase 10.1103/PhysRev.155.1554}
  {\bibfield  {journal} {\bibinfo  {journal} {Physical Review}\ }\textbf
  {\bibinfo {volume} {155}},\ \bibinfo {pages} {1554} (\bibinfo {year}
  {1967})}\BibitemShut {NoStop}%
\bibitem [{\citenamefont {Carroll}(2003)}]{carroll_textbook_2003}%
  \BibitemOpen
  \bibfield  {author} {\bibinfo {author} {\bibfnamefont {S.~M.}\ \bibnamefont
  {Carroll}},\ }\href@noop {} {\emph {\bibinfo {title} {Spacetime and Geometry:
  An Introduction to General Relativity}}}\ (\bibinfo  {publisher} {Pearson},\
  \bibinfo {year} {2003})\BibitemShut {NoStop}%
\bibitem [{\citenamefont {Fletcher}(1960)}]{fletcher_local_1960}%
  \BibitemOpen
  \bibfield  {author} {\bibinfo {author} {\bibfnamefont {J.~G.}\ \bibnamefont
  {Fletcher}},\ }\href {\doibase 10.1103/RevModPhys.32.65} {\bibfield
  {journal} {\bibinfo  {journal} {Reviews of Modern Physics}\ }\textbf
  {\bibinfo {volume} {32}},\ \bibinfo {pages} {65} (\bibinfo {year}
  {1960})}\BibitemShut {NoStop}%
\bibitem [{\citenamefont {Goldberg}(1980)}]{goldberg_invariant_1980}%
  \BibitemOpen
  \bibfield  {author} {\bibinfo {author} {\bibfnamefont {J.~N.}\ \bibnamefont
  {Goldberg}},\ }in\ \href@noop {} {\emph {\bibinfo {booktitle} {General
  {Relativity} and {Gravitation}}}},\ Vol.~\bibinfo {volume} {1}\ (\bibinfo
  {publisher} {Plenum Press},\ \bibinfo {address} {New York},\ \bibinfo {year}
  {1980})\ pp.\ \bibinfo {pages} {469--489}\BibitemShut {NoStop}%
\bibitem [{\citenamefont
  {Kosmann-Schwarzbach}(2011)}]{kosmann-schwarzbach_noether_2011}%
  \BibitemOpen
  \bibfield  {author} {\bibinfo {author} {\bibfnamefont {Y.}~\bibnamefont
  {Kosmann-Schwarzbach}},\ }in\ \href {\doibase 10.1007/978-0-387-87868-3_3}
  {\emph {\bibinfo {booktitle} {The {Noether} {Theorems}: {Invariance} and
  {Conservation} {Laws} in the {Twentieth} {Century}}}},\ \bibinfo {series and
  number} {Sources and {Studies} in the {History} of {Mathematics} and
  {Physical} {Sciences}},\ \bibinfo {editor} {edited by\ \bibinfo {editor}
  {\bibfnamefont {Y.}~\bibnamefont {Kosmann-Schwarzbach}}\ and\ \bibinfo
  {editor} {\bibfnamefont {B.~E.}\ \bibnamefont {Schwarzbach}}}\ (\bibinfo
  {publisher} {Springer New York},\ \bibinfo {address} {New York, NY},\
  \bibinfo {year} {2011})\ pp.\ \bibinfo {pages} {55--64}\BibitemShut {NoStop}%
\end{thebibliography}
\end{document}